\documentstyle[12pt]{article}
\begin{document}
\renewcommand{\thefootnote}{\fnsymbol{footnote}}
\newpage
\pagestyle{empty}
\setcounter{page}{0}



\newcommand{\norm}[1]{{\protect\normalsize{#1}}}
\newcommand{\LAP}
{{\small E}\norm{N}{\large S}{\Large L}{\large A}\norm{P}{\small P}}
\newcommand{\sLAP}{{\scriptsize E}{\footnotesize{N}}{\small S}{\norm
L}$
${\small A}{\footnotesize{P}}{\scriptsize P}}
\begin{minipage}{5.2cm}
\begin{center}
{\bf G{\sc\bf roupe d'} A{\sc\bf nnecy}\\
\ \\
Laboratoire d'Annecy-le-Vieux de Physique des Particules}
\end{center}
\end{minipage}
\hfill
\hfill
\begin{minipage}{4.2cm}
\begin{center}
{\bf G{\sc\bf roupe de} L{\sc\bf yon}\\
\ \\
Ecole Normale Sup\'erieure de Lyon}
\end{center}
\end{minipage}
\centerline{\rule{12cm}{.42mm}}

\vspace{20mm}

\begin{center}
{\bf \Large{The massive spinning particle revisited}}
\\[1cm]

\vspace{10mm}

{\large F. Delduc$^{1a}$, E. Ivanov$^{2b}$ and E.  Sokatchev$^{3c}$}

{\em Laboratoire de Physique Th\'eorique }\LAP\footnote{URA 14-36 du
CNRS, associ\'ee \`a l'Ecole Normale Sup\'erieure de Lyon et \`a
l'Universit\'e de Savoie

\noindent
$^1$ Groupe de Lyon: ENSLAPP, 46, all\'ee d'Italie, 69364 Lyon Cedex 07,
France \\
\noindent
$^2$ Joint Institute for Nuclear Research, 141980 Dubna, Russia \\
\noindent
$^3$ Groupe d'Annecy: LAPP, Chemin de Bellevue BP 110, F-74941
Annecy-le-Vieux Cedex, France. \\
\noindent
$^a$ e-mail address: Francois.Delduc@enslapp.ens-lyon.fr \\
\noindent
$^b$ e-mail address: eivanov@thsun1.jinr.dubna.su \\
\noindent
$^c$ e-mail address: sokatche@lapphp0.in2p3.fr \\
\noindent
}\\

\end{center}
\vspace{20mm}

\centerline{ {\bf Abstract}}
\vskip5mm
\indent
We propose a formulation of the massive spinning particle in terms of
physical bosonic and fermionic fields only. We make use neither of
auxiliary objects of the type of $\gamma_5$ nor of gauge fields. The model
is to be used as a suitable starting point for a forthcoming study of the
rigid superparticle.

\vfill
\rightline{\LAP-A-632/96, JINR E2-96-509}
\rightline{ March 1997}

\newpage
\pagestyle{plain}
\renewcommand{\thefootnote}{\arabic{footnote}}

\newpage\setcounter{page}1

\section{Introduction}

The history of the massive spinning particle is 20 years long. Two
different formulations of the spinning particle have been given in
\cite{Howe} and \cite{Casalbuoni} (see also \cite{Berezin},\cite{Plusch}).
The first of them represents a one-dimensional sigma model with local
world-line supersymmetry (in the formulation of \cite{Casalbuoni} the
world-line supersymmetry is not manifest). The main idea of both approaches
is that the fermionic superpartners $\psi^\mu(t)$ of the target space-time
coordinates $x^\mu(t)$ become gamma matrices upon quantization. The aim is
to interpret one of the first-class constraints of the theory as the Dirac
equation $p^\mu\gamma_\mu\vert\Omega\rangle = m\vert\Omega\rangle$.
However, in the massive case one encounters an obstacle. The left-hand side
of the Dirac equation contains a gamma matrix which corresponds to a
fermion field, whereas the right-hand side only involves the mass. The way
out of this problem is to employ an auxiliary fermion field $\psi_5(t)$
which is identified with the matrix $\gamma_5$ (in a four-dimensional
target space) upon quantization. Then the physical fermion fields
$\psi^\mu(t)$ are interpreted as $\gamma^\mu\gamma_5$ rather than simply
$\gamma^\mu$. Thus one obtains the Dirac equation multiplied by $\gamma_5$:
$p^\mu\gamma_\mu \gamma_5 \vert\Omega\rangle =
m\gamma_5\vert\Omega\rangle$. Notice that besides the auxiliary field
$\psi_5(t)$ the approach of \cite{Howe} involves a set of fields gauging
the world-line $N=1$ superconformal group.

In this paper we propose a new approach in which we make no use of any
auxiliary or gauge fields. One way to arrive at such a formulation is to
eliminate $\psi_5(t)$ and the gauge fields from the model of \cite{Howe}
and obtain a theory involving the physical fields $x^\mu(t)$, $\psi^\mu(t)$
only. In fact, we find a whole one-parameter family of such actions. They
can be formulated in superspace in terms of a single superfield
$X^\mu(t,\theta)$. In this form they possess a manifest world-line $N=1$
conformal supersymmetry. We show that a suitable change of the superfield
variables reduces the entire family of actions to its simplest case. A
characteristic feature of the corresponding component action is the
presence of a transverse projection operator made out of the bosonic
velocities in the kinetic term of the fermions. This implies that only
$D-1$ of the $D$ fermion fields $\psi^\mu(t)$ ($D$ is the dimension of the
target space-time) are effectively present, the last one being gauged away
by the local supersymmetry. We comment on some generic problems in the
straightforward application of the standard Dirac quantization procedure.
To successfully quantize the model, we first fix a local supersymmetry
gauge by introducing a Lagrange multiplier term in the action. Only after
this it becomes possible to carry out the Dirac procedure. We show that the
$D-1$ transverse fermions can be realized as the space-like gamma matrices
$\gamma^i$ ($i= 1,2,\ldots, D-1 $). Thus the wave function of the quantized
theory is a spinor of the rotation group $SO(D-1)$. A first-class
constraint gives this spinor the mass $m$ and we obtain a spin one half
massive particle. The hidden Lorentz covariance of the theory is manifested
in the possibility to construct out of the fields the generators of the
Lorentz group in $D$ dimensions, $SO(1,D-1)$. Then the fact that the wave
function carries spin one half can be verified in a Lorentz- covariant way
by computing the eigenvalue of the Poincar\'e group spin Casimir operator.
For $D$ even the wave function carries an irreducible spin one half
representation, while for $D$ odd it comprises two such representations. In
the latter case our model describes two spin half particles of the same
mass. The final results obtained in this way are equivalent to those in the
$\gamma_5$ approach, although the procedure is quite different.

Our main motivation in addressing this problem was to find a suitable
starting point for the study of the so-called rigid superparticle. The
rigid particle \cite{rigid} is a generalization of the massive particle
which employs a higher order invariant (curvature) as a Lagrangian and
possesses a gauge symmetry of the $W$ type \cite{Ramos}. It is natural to
look for a supersymmetric generalization of the theory based on the massive
spinning particle. This lead us to seek the most economical formulation of
the massive spinning particle. The choice of a simple starting point is
essential for the analysis of the rather complicated gauge invariance of
the rigid superparticle. Moreover, a number of features of the formalism
developed here can immediately be generalized to the rigid superparticle.
Details on this will be presented in a future publication.

It should be mentioned that the elimination of the $\gamma_5$-like variable
and of the gauge fields has also been done (although in a different way)
in ref. \cite{Gauntlett}. The result obtained there coincides with one
member of our series of actions above, but its Hamiltonian analysis has not
been performed.

In section 2 we formulate the  massive spinning particle in terms of a
single superfield $X^\mu(t,\theta)$. We find a one-parameter family of such
actions, all possessing world-line $N=1$ conformal supersymmetry. Then we
discuss the component form of those actions and prove their equivalence. In
section 3 we perform the Dirac quantization of the theory and show that it
describes a massive particle of spin one half. We discuss the cases of even
and odd target-space dimension.

\section{The massive spinning particle action}

\subsection{The superfield action}

As mentioned in the Introduction, one can arrive at the new formulation of the
massive spinning particle by eliminating the auxiliary variables from the
action of ref. \cite{Howe}. In terms of component fields the latter reads:
  \begin{equation}\label{comphowe}
  S={1\over 2}\int dt\; \left[e^{-1} \dot x^2 + m^2 e - \psi\cdot\dot\psi +
  \psi_5\dot\psi_5 + \chi(m\psi_5 - e^{-1}\psi\cdot\dot x) \right]  \ .
  \end{equation}
Here $x^\mu$ and $\psi^\mu$ are the physical bosonic and fermionic fields,
$e$ and $\chi$ are the gauge fields for world-line diffeomorphisms and
local supersymmetry (einbein and gravitino). The symbol $\cdot$ stands for
a contraction of two space-time vector indices, e.g. $\psi\cdot\dot x
\equiv \psi^\mu\dot x_\mu$. Note that despite the presence of a kinetic
term for $\psi_5$ it is an auxiliary fermion, since the field equation of
$\chi$ allows to express $\psi_5$ in terms of other fields. The action
(\ref{comphowe}) can be put in a superfield form as follows \cite{Ramos}:
  \begin{equation}\label{sfhowe}
  S={1\over 2}\int dtd\theta\; \left[ E^{-1} DX\cdot \dot X - BDB + 2m BE^{1/2}
  \right] \ .
  \end{equation}
Here one uses the even superfields $X^\mu=x^\mu(t) + \theta\psi^\mu(t)$, $E
= e(t) + \theta \chi(t)$ and the odd one $B = \psi_5(t) + \theta b(t)$. The
auxiliary bosonic field $b(t)$ has been eliminated in (\ref{comphowe}). The
spinor derivative is defined as $D=\partial_\theta + \theta\partial_t$. The
idea now is to eliminate both the auxiliary ($B$) and gauge ($E$)
superfields using their field equations:
  \begin{eqnarray}
  {\delta\over\delta E}: && -E^{-2}DX\cdot \dot X + m BE^{-1/2} = 0 \quad
  \rightarrow \quad B= {1\over m} E^{-3/2} DX\cdot \dot X \nonumber \\
  {\delta\over\delta B}: &&  -2DB +2m E^{1/2} = 0 \quad
  \rightarrow  \quad E = {1\over m} \sqrt{\dot X^2 - DX\cdot D\dot X} +
  O(DX\cdot \dot X )  \ .   \nonumber
  \end{eqnarray}
Replacing $B$ and $E$ in (\ref{sfhowe}) one obtains an action in terms of
the physical superfield $X^\mu$ only:
  \begin{equation}\label{act1}
S =  m\int dtd\theta\;{DX\cdot \dot X \over \sqrt{\dot X^2 - DX\cdot D\dot X}}
\ .
  \end{equation}
The procedure of obtaining (\ref{act1}) from (\ref{sfhowe}) guarantees that
the two actions are classically equivalent. As we shall show below, the
quantization of (\ref{act1}) and (\ref{sfhowe}) goes along different lines,
but leads to equivalent results.

Even though the action (\ref{act1}) does not contain any gauge objects, it
still has manifest local supersymmetry ($N=1$ conformal world-line
supersymmetry). It is realized in the form of superdiffeomorphisms
  \begin{equation}\label{diff}
  \delta t = \lambda(t,\theta)\; , \quad \delta\theta = \eta(t,\theta)\ ,
  \end{equation}
where the parameters satisfy the constraint $D\lambda = \eta + \theta
D\eta$. In components this implies $\lambda = \rho(t) + \theta\epsilon(t)$,
$\eta = \epsilon(t) + {1\over 2} \theta \dot\rho(t)$, where $\rho(t)$ and
$\epsilon(t)$ are the parameters of world-line diffeomorphisms and local
supersymmetry, correspondingly. From (\ref{diff}) one obtains the
transformation laws
  $$
  \delta X =0\;, \quad \delta D = -(D\eta)D\;, \quad \delta \partial_t =
  -2(D\eta) \partial_t - \dot\eta D \;, \quad \delta (dtd\theta) =
  (D\eta)dtd\theta \ .
  $$
With this it is easy to verify that the action (\ref{act1}) is invariant.
Indeed, one finds: $\delta(DX\cdot \dot X) = -3(D\eta) DX\cdot \dot X$ and
$\delta(\dot X^2 - DX\cdot D\dot X) = -4(D\eta)(\dot X^2 - DX\cdot D\dot X)
- 3\dot\eta DX\cdot \dot X$, so the Lagrangian in (\ref{act1}) transforms
with the weight $-(D\eta)$ opposite to that of the superspace measure
$dtd\theta$.

The action (\ref{act1}) has an obvious generalization:
  \begin{equation}\label{act2}
  S =  m\int dtd\theta\;{DX\cdot \dot X \over \sqrt{\dot X^2 +\alpha DX\cdot
  D\dot X}}
  \end{equation}
where $\alpha$ is a parameter (later on we shall show that $\alpha\neq-2$).
The $N=1$ conformal world-line supersymmetry of (\ref{act1}) is preserved
by (\ref{act2}). Although not immediately obvious, all of these actions are
equivalent up to a change of variables. To see this, make an infinitesimal
shift of the parameter $\alpha$,
$\alpha \rightarrow \alpha + \delta \alpha$. The corresponding variation
of the superfield Lagrangian is
  \begin{equation}\label{shift}
  \delta L = -{\delta \alpha \over 2} {DX\cdot D\dot X\; DX\cdot \dot X \over
  (\dot X^2 +\alpha DX\cdot D\dot X)^{3/2}}\;.
  \end{equation}
This variation
can be compensated by an appropriate variation of the superfield $X^\mu$.
Consider the following variation of $X^\mu$
  \begin{equation} \label{chvariab}
  \delta X^\mu  = f(X)\; DX\cdot \dot X\; DX^\mu\;,
  \end{equation}
$f(X)$ being a not yet specified scalar function of $X$. It is easy to
find the variation of $L$ under this transformation
  \begin{equation} \label{varL}
  \delta L = - f
  {(\alpha + 2)\dot X^2 +\alpha DX\cdot D\dot X\over
  (\dot X^2 +\alpha DX\cdot D\dot X)^{3/2}}
  DX\cdot D\dot X\; DX\cdot \dot X\;.
  \end{equation}
Choosing
$$
f =  -{\delta \alpha \over 2}
\left((\alpha + 2)\dot X^2 +\alpha DX\cdot D\dot X \right)^{-1}
$$
(it is not singular if $\alpha \neq -2$) we see that the variations
(\ref{shift}) and (\ref{varL}) cancel. Thus any shift of $\alpha$ can be
compensated by a suitable redefinition (\ref{chvariab}) of $X^\mu$, which
proves the classical equivalence of the actions (\ref{act2}) with different
$\alpha$'s except the special case $\alpha = -2$. \footnote{Interestingly
enough, the same argument can be repeated if the constant parameter
$\alpha$ is replaced by a {\it superfield} $\Phi(t,\theta)$. Then the shift
of $\alpha$ is upgraded to a full gauge invariance, the superfield $\Phi$
being pure gauge.}

The established equivalence of all actions (\ref{act2}) allows us to
concentrate on the simplest case $\alpha=0$:
  \begin{equation}\label{alpha0}
  S =  m\int dtd\theta\;{DX\cdot \dot X \over \sqrt{\dot X^2}}
  \end{equation}
Thus we avoid the non-linearity in the fermion term $DX\cdot D\dot X$ and
the appearance of second-order time derivatives of $x^\mu$ in the component
action.

In conclusion we note that (\ref{act2}) can also be written down in the form
  \begin{equation}\label{act2al}
  S = \int dtd\theta\;E_\alpha^{-1} DX\cdot \dot X \;,
  \end{equation}
where
  $$
  E_\alpha = {1\over m} \sqrt{\dot{X}^2 +\alpha DX\cdot D \dot X}
  \left[ 1 + \frac{2-\alpha}{4} \frac{ DX\cdot \dot X  (2 D\dot X\cdot \dot X +
  (1-\alpha) DX\cdot \ddot X)} {( \dot{X}^2 + \alpha DX\cdot D \dot X )^2}
\right]
  $$
is a composite supervielbein with the transformation law $\delta E_\alpha =
-2 (D \eta) E_\alpha$. For $\alpha = -1$ this $E_\alpha$ is just the one
obtained by solving the equations for the auxiliary superfields $E$ and $B$
following from the action (\ref{sfhowe}). We do not know whether the
generalized action for any $\alpha$ (\ref{act2al}) can be obtained from some
analog of (\ref{sfhowe}) by a similar procedure.

\subsection{The component action} \label{comp}

The component form of the action (\ref{alpha0}) is very easy to obtain:
  \begin{equation}\label{action}
  L =  m\sqrt{\dot x^2}  + \dot\psi^\mu\left(\delta_{\mu\nu} -
  {\dot x_\mu \dot x_\nu \over\dot x^2} \right) \psi^\nu \ .
  \end{equation}
Here we have made the field redefinition $\psi \rightarrow (\dot
x^2)^{1/4}\psi$ in order to have a fermion field with vanishing weight
under diffeomorphisms. Since the action $S=\int dt L$ with $L$ from
(\ref{action}) is obtained from the superspace action (\ref{alpha0}), it is
automatically invariant under the local supersymmetry transformations
obtained from (\ref{diff}) by taking into account the filed redefinition
above:
  \begin{equation}\label{locsusy}
  \delta x^\mu = \epsilon\psi^\mu\ ,\quad
  \delta \psi^\mu = \epsilon {\dot x^\mu\over\sqrt{\dot x^2}}
  -{1\over 2{\dot x^2}}(\dot\epsilon\psi\cdot\dot x+\epsilon\dot\psi\cdot\dot
  x)\psi^\mu\ .
  \end{equation}

 The component Lagrangian (\ref{action}) allows a
direct comparison with the original action (\ref{comphowe}) of ref.
\cite{Howe}. The presence of the projection operator $\delta_{\mu\nu} -
\dot x_\mu \dot x_\nu /\dot x^2$ in the fermionic kinetic term in
(\ref{action}) suggests to split the fermion field as follows:
 $$
\psi^\mu = \psi^\mu_\perp  + \dot x^\mu\psi_\parallel \ ,
 $$
where the projection $\psi^\mu_\perp \equiv \left(\delta^{\mu\nu} - {\dot
x^\mu \dot x^\nu /\dot x^2} \right)\psi_\nu $ satisfies the
orthogonality condition $\dot x\cdot\psi_\perp = 0$. We can incorporate this
condition in the Lagrangian with a Lagrange multiplier. Then (\ref{action})
can be rewritten as follows:
  \begin{equation}\label{rewrite}
L =  m\sqrt{\dot x^2}  + \dot\psi_\perp\cdot(\psi_\perp + \dot x
\psi_\parallel) + \lambda\dot x\cdot\psi_\perp \ .
\end{equation}
Now, take the Lagrangian in (\ref{comphowe}) and eliminate the einbein
field $e$ from it. The result is
  \begin{equation}\label{eliminate}
L = m\sqrt{\dot x^2} - {1\over 2}\psi\cdot\dot\psi + {1\over
2}\psi_5\dot\psi_5 - {1\over 2} m\chi ({\psi\cdot\dot x\over \sqrt{\dot
x^2}} - \psi_5 ) \ .
\end{equation}
Then one easily identifies the two Lagrangians (\ref{rewrite}) and
(\ref{eliminate}) by making the change of variables $\psi =
\sqrt{2}\psi_\perp + {\dot x\over\sqrt{2}} \psi_\parallel$, $\psi_5 =
{\sqrt{\dot x^2} \over\sqrt{2}}
\psi_\parallel$, $\chi = - {\sqrt{2}\over m}\sqrt{\dot x^2}\lambda$.
Thus, we have given a second proof of the classical equivalence of the two
description of the spinning particle.

\section{Quantization}

\subsection{Hamiltonian analysis}

The Lagrangian (\ref{action}) has an unusual feature, the presence of the
projection operator $\delta_{\mu\nu} - \dot x_\mu \dot x_\nu /\dot x^2$ in
the fermionic kinetic term. This makes the direct application of the
standard Dirac quantization procedure for constrained systems \cite{Dirac}
surprisingly difficult. Attempting to define the momenta conjugate to $x$
and $\psi$ directly from (\ref{action}) does not allow us to find the
primary constraints. The reason is the presence of a fermionic velocity
$\dot\psi$ in the bosonic momentum conjugate to $x$. The fermionic momentum
conjugate to $\psi$ is totally degenerate, i.e. none of the fermionic
velocities can be expressed through phase space variables. Therefore we
find no way to eliminate $\dot\psi$ from the momentum conjugate to $x$ and
to obtain a meaningful primary constraint (the superanalog of $p^2=m^2$ in
the purely bosonic case). We tried to solve this problem by using a
``first-order formalism", i.e. introducing a new bosonic variable
$q^\mu=\dot x^\mu$ in order not to have bosonic velocities mixed up with
the fermionic ones. Unfortunately, this trick creates a new unexpected
difficulty. When stabilizing the primary bosonic constraint with the total
Hamiltonian, we obtained an equation containing a Lagrange multiplier
(gauge field) {\it multiplied by a fermion}. Such an equation neither is a
secondary constraint (because it contains a Lagrange multiplier) nor allows
to determine this Lagrange multiplier (because its coefficient is odd, i.e.
non-invertible). This is a rather peculiar situation, in which the standard
Dirac method fails. \footnote{We would like to mention that a similar
problem occurs in ref. \cite{Gomis}, where an attempt to quantize the
action of ref. \cite{Gauntlett} has been made. In our opinion, the authors
of ref. \cite{Gomis} have not found a satisfactory way out.}

We can propose two ways out. One is to simply say that our theory
(\ref{action}) is classically equivalent to the original one
(\ref{comphowe}), therefore they should have the same spectrum. If,
however, one wants to have an independent quantization of (\ref{action}),
we propose to make use of the local supersymmetry (\ref{locsusy}) of our
Lagrangian (\ref{action}). Note that the first term in $\delta\psi^\mu$ is a
pure shift by the supersymmetry parameter $\epsilon$ with no a time
derivative on it. This implies the existence of the ``perfect"
(i.e. globally achievable) gauge
  \begin{equation}\label{gaguge}
\dot x\cdot \psi=0 \ .
\end{equation}
Such a gauge can be incorporated into the Lagrangian with a Lagrange
multiplier, after which we obtain a much simpler Lagrangian:
  \begin{equation}\label{gaugefix}
L =  m\sqrt{\dot x^2}  + \dot\psi\cdot\psi + \lambda\dot x\cdot \psi \ .
\end{equation}
Note that we have absorbed the troublesome mixed terms containing fermionic
and bosonic velocities into the Lagrange multiplier $\lambda$. In is not
hard to show that the Lagrangian (\ref{gaugefix}) gives rise to the same
field equations as (\ref{action}) with the gauge (\ref{gaguge}) imposed by
hand and on shell. The reason for this is the ``perfect" nature of the gauge
(\ref{gaguge}). This would not be true, for instance, if we tried to
incorporate a gauge like $\dot x^2=1$ into the Lagrangian, because it is
achieved through a time derivative of the diffeomorphism parameter.

The Hamiltonian analysis of (\ref{gaugefix}) is straightforward. The
momentum conjugate to $x^\mu$ is
 $$
p^\mu = m{\dot x^\mu\over \sqrt{\dot x^2}} + \lambda\psi^\mu \ ,
 $$
which gives rise to the bosonic primary constraint
  \begin{equation}\label{bcon}
p^2-m^2-2m\lambda p\cdot\psi \approx 0 \ .
\end{equation}
Further, we obtain two fermionic primary constraints:
  $$
p^\psi_\mu-\psi_\mu \approx 0\;, \quad p^\lambda \approx 0 \ .
  $$
The canonical Hamiltonian vanishes and the total one is
 $$
H_T = e(p^2-m^2-2m\lambda p\cdot\psi) + \sigma\cdot(p^\psi-\psi) + \omega
p^\lambda
\ ,
 $$
where $e,\sigma,\omega$ are Lagrange multipliers. The stabilization of the
fermionic constraint $p^\lambda \approx 0$ generates a secondary one,
  \begin{equation}\label{seccon}
p\cdot\psi \approx 0
\end{equation}
and that of $p^\psi_\mu-\psi_\mu \approx 0$ leads to an expression for
the Lagrange
multiplier $\sigma^\mu = -mep^\mu \lambda$. The bosonic constraint
(\ref{bcon}) then
turns out stable. Further, the stabilization of the secondary constraint
(\ref{seccon}) produces a new one, $\lambda\approx 0$, whose stabilization
in turn fixes the Lagrange multiplier $\omega=0$. The only remaining
Lagrange multiplier $e$ is the gauge field for the diffeomorphisms.

The constraints $p^\lambda $ and $\lambda$ form a trivial second class pair
and after introducing the appropriate Dirac bracket they drop out. The
remaining constraints $p^2-m^2$, $\beta_\mu \equiv
p^\psi_\mu-\psi_\mu$ and $\gamma\equiv p\cdot\psi$ satisfy the algebra (only
the non-vanishing brackets are shown)
 $$
[\beta_\mu,\beta_\nu]_+ = -2\delta_{\mu\nu} \;, \quad [\beta_\mu,\gamma]_+ =
p_\mu \ .
 $$
Now we split $\beta_\mu$ into $\beta_\mu^\perp = (\Pi\beta)_\mu$ and
$\beta^\parallel = p\cdot\beta$ (where $\Pi_{\mu\nu} =
\delta_{\mu\nu}-p_\mu p_\nu / p^2$ is the projector orthogonal to $p$) and
replace $\beta^\parallel$ by the linear combination $\hat\beta =
\beta^\parallel + \gamma$. After this the algebra becomes
 $$
[\beta_\mu^\perp,\beta_\nu^\perp]_+ = -2\Pi_{\mu\nu} \;, \quad
[\hat\beta,\gamma]_+ = m^2 \ .
 $$
Thus we see that $\hat\beta$ and $\gamma$ form a second class pair. The
same is true for the orthogonal projections $\beta_\mu^\perp$, since the
projector $\Pi$ is invertible in the subspace of vectors orthogonal to $p$.
Then we introduce the Dirac bracket
 $$
[A,B]^* = [A,B] + {1\over 2} [A,\beta_\mu^\perp] \Pi^{\mu\nu}
[\beta_\nu^\perp,B] -{1\over m^2} [A,\hat\beta][\gamma,B] -{1\over m^2}
[A,\gamma][\hat\beta,B]
 $$
and impose the second class constraints $\beta_\mu^\perp,\hat\beta,\gamma$
strongly. The only remaining first class constraint is the mass-shell
condition $p^2-m^2=0$ which is imposed on the quantum states. Finally, it is
easy to find the Dirac bracket of the fermionic variables $\psi^\perp =
\Pi\psi$:
  \begin{equation}\label{gamma}
[\psi_\mu^\perp,\psi_\nu^\perp]^*_+ = {1\over 2} \Pi_{\mu\nu} \ .
\end{equation}
Note that the parallel projection $\psi^\parallel = p\cdot\psi$ vanishes,
since it coincides with the second class constraint $\gamma$, which is now
imposed strongly.

\subsection{Quantization and interpretation of the results}

The form of the Dirac bracket (\ref{gamma}) suggests to interpret the
fermionic variables $\psi^\perp_\mu$ after quantization as gamma matrices.
However, we should
take account of the projection operator in (\ref{gamma}). Its r\^ole is to
project out a $D-1$-dimensional ``transverse'' subspace of the space spanned
by the $D$-dimensional vectors. This can be done, for instance, by going to
the rest frame in which $p_\mu=(m,0,\ldots,0)$. Then $\Pi$ just
projects out the space-like part of a vector. Thus we see that a
possible realization of the quantum fermionic variables is $\psi^\perp_\mu
\ \rightarrow \ 1/2 \gamma_i$ where $i=1,\ldots,D-1$ and $\{\gamma_i,\gamma_j\}
= 2\delta_{ij}$. Such a realization can be achieved on quantum states
$\vert\Omega\rangle$ representing spinors of the minimal required
dimension. If $D=2n$, the dimension of the spinor is $2^{n-1}$; if
$D=2n+1$, it is $2^n$. We conclude that our model describes a spinor
particle of mass $m$.

Let us compare this result to the quantization of the spinning particle in
the form (\ref{comphowe}) of ref. \cite{Howe}. There one has $D$ fermionic
variables $\psi_\mu$ and an additional one $\psi_5$. It is not hard to show
that their Dirac brackets are equivalent to the anticommutators of the
gamma matrices $\gamma_\mu$ (or, which is the same, of
$\gamma_\mu\gamma_5$) and $\gamma_5$ (clearly, this interpretation makes
sense if $D=2n$, where $\gamma_5 =
\epsilon^{\mu_1\ldots\mu_D} \gamma_{\mu_1}\ldots\gamma_{\mu_D}$ exists as an
independent matrix). Further, one of the first-class constraint of the theory
takes the form of the Dirac equation multiplied by $\gamma_5$:
  \begin{equation}\label{diraceq}
  p^\mu\gamma_\mu\gamma_5\vert\Omega\rangle = m\gamma_5\vert\Omega\rangle \
  .
  \end{equation}
In the rest frame $p_\mu=(m,0,\ldots,0)$ eq. (\ref{diraceq})
becomes
  \begin{equation}\label{eq}
  (\gamma_0-1)\gamma_5\vert\Omega\rangle=0 \ .
  \end{equation}
The matrix $\gamma_0$ in $D$ dimensions can be realized as follows:
$(\gamma_0)_D =(\gamma_0)_{D=2}\otimes I$ where $(\gamma_0)_{D=2} =
\left(\begin{array}{rr} 1&0\\ 0&-1 \end{array}\right)$ is $\gamma_0$ in two
dimensions and $I$ is the identity matrix for the rest of the spinor space.
Then equation (\ref{eq}) serves as a projection condition that selects one
half of the spinor state $\vert\Omega\rangle$. Thus, in the case of even
target space dimension $D=2n$ the spinors of dimension $2^n$
are cut down to dimension $2^{n-1}$. Note that our model leads to the same
type of quantum state.

In the case of odd target space dimension $D=2n+1$ there is no natural
object of the type of $\gamma_5$. Nevertheless, one could apply the same
quantization scheme to the model of ref. \cite{Howe} as follows. One goes
to the next even dimension $D=2n+2$ and there realizes the fermion
variables $\psi_\mu$ as the first $2n+1$ of the $2n+2$ matrices
$\gamma_\mu\gamma_5$ and the variable $\psi_5$ as the matrix $\gamma_5$ of
that even-dimensional space. Then one can write down the equation
(\ref{diraceq}) which will cut the spinors from dimension $2^{n+1}$ down to
dimension $2^n$. However, these spinors will still be twice as big as the
minimal required size in an odd-dimensional target space with $D=2n+1$.
Thus, the quantum state obtained in this case is reducible and consists of
two massive spinning particles (see also \cite{Plusch}). Exactly the same
happens in the model proposed in this paper: In the odd-dimensional case we
have found spinors of dimension $2^n$ and not $2^{n-1}$. We conclude that
although the two models look quite different, the final results are in fact
equivalent. This is not surprising in view of the classical equivalence of
the two actions exhibited in section 2.

In the argument above we made use of the non-covariant rest frame. However,
the quantum spectrum of our model can be determined in a Lorentz covariant
manner as well. In order to show that the theory describes a spin 1/2 state
(or its generalization to $D$ dimensions) it is sufficient to compute the
eigenvalue(s) of the Casimir operator(s) of the Poincar\'e group. We recall
that these operators (besides the mass Casimir $P^2$) are constructed
following the example of the Pauli-Lubanski vector in four dimensions $W_\mu =
{1\over 2}\epsilon_{\mu\nu\lambda\rho}P^\nu L^{\lambda\rho}$ where
$L_{\mu\nu}$ is the generator of the Lorentz group. In the general case one
can form a set of tensors $W_{\mu_1\ldots\mu_{D-2k-1}} =
\epsilon_{\mu_1\ldots\mu_{D-2k-1}
\nu\lambda_1\lambda_2\ldots \lambda_{2k-1}\lambda_{2k}} P^\nu
L^{\lambda_1\lambda_2} \ldots L^{\lambda_{2k-1}\lambda_{2k}}$. Their
squares $W^2_{\mu_1\ldots\mu_{D-2k-1}}$ are Poincar\'e invariants (except
for the case $2k+1=D$ where the scalar $W$ is invariant by
itself).\footnote{In the rest frame these operators are reduced to the
Casimir operators of the little group $SO(D-1)$.} In order to compute the
eigenvalues of these operators we need a realization of the Lorentz
generator in terms of the coordinates and momenta of our model. It is given
by the Noether charges corresponding to Lorentz transformations:
  \begin{equation}\label{gener}
  J_{\mu\nu} = x_\mu p_\nu - x_\nu p_\mu -  \psi^\perp_\mu \psi^\perp_\nu
   + \psi^\perp_\nu \psi^\perp_\mu \ .
  \end{equation}
Despite the apparent presence of transverse fermions only in (\ref{gener})
it is not hard to show that the Dirac bracket of these charges reproduces
the Lorentz algebra:
  $$
  [J_{\mu\nu}, J_{\lambda\rho}]^{*} = \delta_{\mu\lambda}J_{\nu\rho} -
  \delta_{\mu\rho}J_{\nu\lambda} - \delta_{\nu\lambda}J_{\mu\rho} +
  \delta_{\nu\rho}J_{\mu\lambda} \ .
  $$
The eigenvalue computation will be illustrated in the four-dimensional
case. The spin Casimir operator is $W_\mu^2 = -{1\over 2} P^2 L^{\mu\nu}
L_{\mu\nu} + (P^\mu L_{\mu\nu})^2$. Inserting into it $L_{\mu\nu}$ in the
form (\ref{gener}), noticing that the $xP$ part of it drops out, using the
transversality condition $P^\mu \psi^\perp_\mu\vert\Omega\rangle = 0$ ($P$
and $\psi^\perp$ are now operators; $P_\mu \sim p_\mu$) and the
anticommutator relation (\ref{gamma}) we easily find $W^2\vert\Omega\rangle
= {3\over 4}m^2\vert\Omega\rangle$ which is the eigenvalue corresponding to
spin one half. Note that this computation resembles the one made using the
gamma matrix realization of the Lorentz generator $L_{\mu\nu} = x_\mu P_\nu
- x_\nu P_\mu - {1\over 4}(\gamma_\mu
\gamma_\nu - \gamma_\nu \gamma_\mu)$ and the Dirac equation (\ref{diraceq})
in its covariant form. The higher-order invariants in the case $D>4$ can be
computed in a similar way. Knowing the set of eigenvalues of the Casimir
operators, we can determine the type of Poincar\'e group irrep described by
our particle model, but we cannot tell if it is degenerate (as was the case
of odd space-time dimension above).

\vspace{0.5cm}

\noindent{\Large\bf Acknowledgement}
\vspace{0.3cm}

\noindent E.I. thanks A. Pashnev and E.S. thanks P. Howe for useful
discussions. The authors are grateful to M.S. Plyushchay for a number of
useful comments and for turning our attention to some references. The work
of E.I. was partially supported by the Russian Foundation of Basic Research
grants RFBR 96-02-17634 and RFBR-DFG 96-02-00180G, by the INTAS grant
INTAS-94-2317 and by a grant of the Dutch NWO organization.

\end{document}